\documentclass[journal,10pt]{IEEEtran}
\pdfoutput=1
\usepackage{cite}

\usepackage{amsmath}
\usepackage{algorithm}
\usepackage{algorithmic}
\usepackage{subfigure}
\usepackage{amssymb}
\usepackage{graphicx}
\usepackage{empheq}

\usepackage{array}
\usepackage{mathtools}
\usepackage{graphicx}
\usepackage{cases}
\usepackage[colorinlistoftodos]{todonotes}
\usepackage[colorlinks=true, linkcolor=black,citecolor=black,urlcolor=black]{hyperref}
\usepackage{indentfirst}
\usepackage{multirow}
\usepackage{booktabs}
\usepackage{makecell}
\allowdisplaybreaks[4]

\usepackage{url}
\usepackage[hyphenbreaks]{breakurl}

\begin{document}

% \linenumbers

\title{\huge Towards Routing and Edge Computing in Satellite-Terrestrial Networks: A Column Generation Approach}

\author{
Yuan~Liao, Kan~Cheng, Fan~Lu, Hao~Jin and Zhaohui~Yang, \IEEEmembership{Member,~IEEE}

\thanks{Yuan Liao, Kan Cheng and Hao Jin are with the Institute of Remote Sensing Satellite, China Academy of Space Technology, Beijing 100048, China (e-mail: yuanliao3435@outlook.com; ckjack\_007@163.com; 2434790063@qq.com).

Fan Lu is with the Institute of Spacecraft System Engineering, China Academy of Space Technology, Beijing 100048, China (e-mail: fan8736@pku.edu.cn).

Zhaohui Yang is with the College of Information Science and Electronic Engineering, Zhejiang University, and also with Zhejiang Provincial Key Laboratory of Info. Proc., Commun. \& Netw. (IPCAN), Hangzhou, 310027, China (e-mail:yang\_zhaohui@zju.edu.cn). 
}
}

\maketitle

\begin{abstract}
Edge computing that enables satellites to process raw data locally is expected to bring further timeliness and flexibility to satellite-terrestrial networks (STNs). In this letter, we propose a three-layer edge computing protocol, where raw data collected by the satellites can be processed locally, or transmitted to other satellites or the ground station via multi-hop routing for further processing. The overall computing capacity of the proposed framework is maximized by determining the offloading strategy and routing formation, subject to channel capacity and hop constraints. Given that the problem scale grows exponentially with the number of satellites and maximum-allowed hops, the column generation approach is employed to obtain the global optimal solution by activating only a subset of variables. Numerical results reveal that the proposed three-layer computing protocol, when tolerating a 5-hop routing latency, achieves a 60\% improvement in computation capacity compared to the single-layer local computing configuration. 
\end{abstract}


\IEEEpeerreviewmaketitle

\section{Introduction}
\label{introduction}

Satellite-terrestrial networks (STNs) are inevitably required to process a massive number of delay-sensitive tasks, e.g., processing image data captured by remote sensing satellites. Although advancements in mobile edge computing (MEC) enable satellites to process raw data onboard, the heterogeneous distribution of computing demands and the limited computing resources on satellites continue to pose significant challenges for STNs \cite{xie2020satellite}. Fortunately, the recent development of satellite mega-constellations and laser inter-satellite communications is expected to overcome these issues, i.e., satellites with heavy data loads can offload tasks to other satellites or ground stations via high-capacity laser links, thereby enhancing the computing capacity of the STN in the upcoming sixth generation (6G) era \cite{bhattacharjee2024demand}.

To better utilize the limited computational resources of STNs, task offloading is one of the most crucial considerations in satellite-assisted MEC systems \cite{xie2020satellite}. In \cite{wu2024multi}, a reinforcement learning-based offloading strategy is employed to determine whether a task should be processed by users or satellites, aiming to minimize system delay and energy consumption. In addition to the task offloading strategy, resource allocation is taken jointly into account to further optimize the system performance in \cite{zhang2023satellite, cao2022edge, zhang2022aerial}. Furthermore, with the rapid development of laser communication technology, the routing problem based on optical laser inter-satellite links (ISLs) has received increasing attention in recent years. The authors of \cite{chen2021analysis} demonstrate that the routing in STNs will become increasingly complex with large-scale constellations. In \cite{wang2023optimization}, the scheduling of dynamic laser ISLs and routing formation are optimized jointly to conserve energy and communication delay in low-earth-orbit (LEO) constellations. To further improve the energy efficiency, the authors of \cite{bhattacharjee2024demand} propose an on-demand routing strategy for LEO constellations with laser ISLs, where the laser links would be activated only when data is transmitted and remain inactive at other times to reduce energy consumption. Most studies on satellite-assisted MEC assume that satellites are not interconnected \cite{wu2024multi,zhang2023satellite, cao2022edge,zhang2022aerial}, with few studies acknowledging that inter-satellite routing could potentially improve the system performance. In \cite{wang2022cdmr}, a novel paradigm is proposed where satellites could compute tasks during the routing process. In \cite{cao2023computing}, a genetic algorithm-based routing and task offloading strategy is proposed to decrease the overall delay in LEO networks. 

\begin{figure}[!t]
\centering
\includegraphics[width=.8\linewidth]{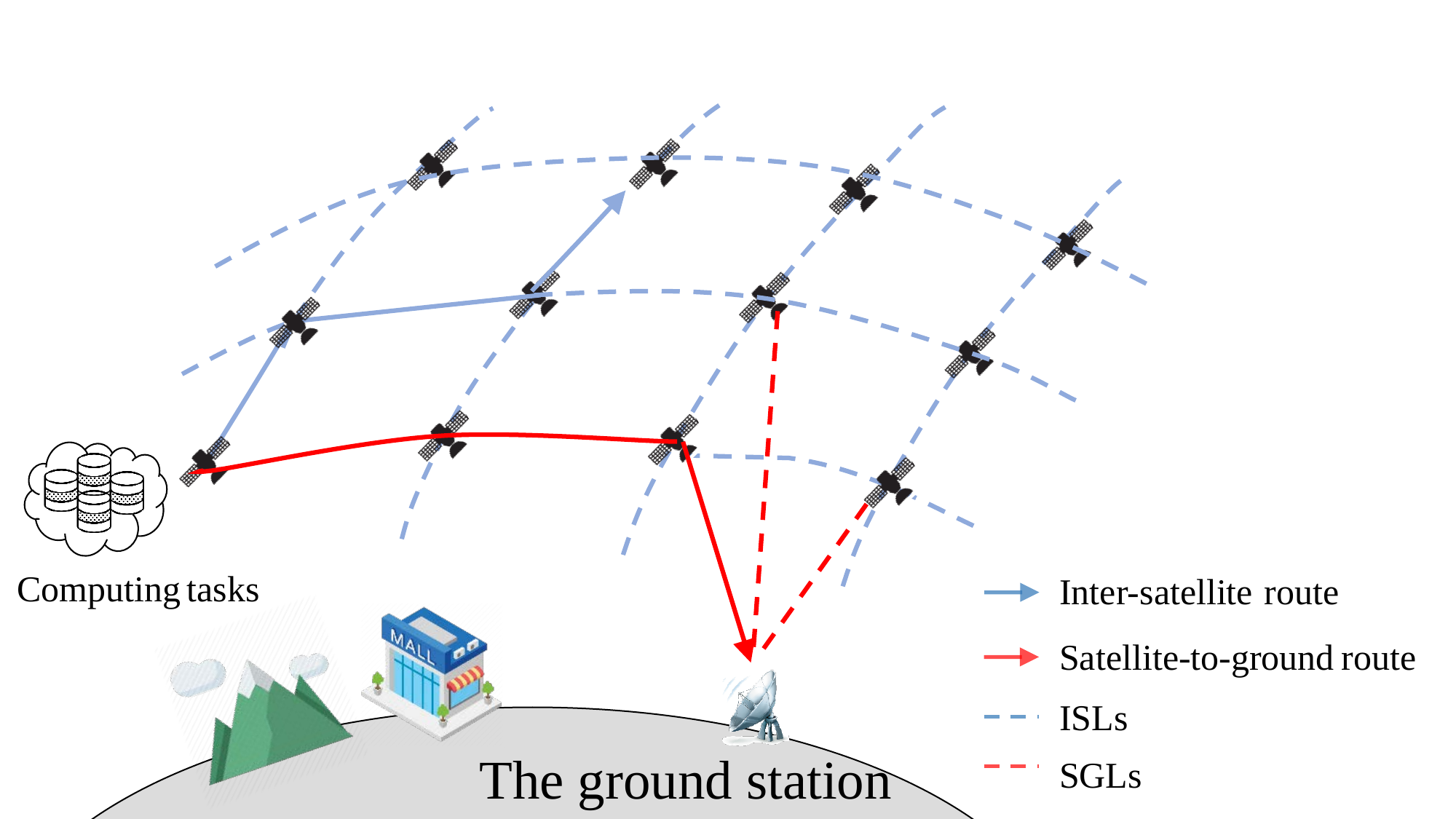}
\caption{Illustration of the STN.}
\label{toy_example}
\end{figure}

This letter presents a three-layer MEC protocol for STNs, where the tasks generated by the satellites can either be processed locally, or transmitted to other satellites or the ground station via optical laser links for further computing.  Specifically, the contributions of this letter are summarized as follows. Firstly, the previous work \cite{cao2023computing} investigates the routing and task offloading within the satellite constellation, while satellites cannot process all row data due to the limited computing resources onboard. With this in mind, we incorporate the ground station with massive computation resources as a cloud terminal. Secondly, instead of optimizing the system delay in \cite{wu2024multi,zhang2023satellite, cao2022edge,zhang2022aerial,cao2023computing}, we focus on maximizing the computing capacity of STNs subject to the maximum hop constraints, guaranteeing that the routing formation satisfies the latency requirement. Thirdly, a linear programming (LP) is formulated to maximize the computation capacity by determining the optimal offloading and multi-hop routing formation for computation-intensive data. Given that the scale of the LP problem grows exponentially with the number of satellites and hops, the column generation method is employed to achieve the global optimal solution by activating only a subset of variables. Simulation results show that the proposed three-layer edge computing protocol could increase the computing capacity by 60\% comparing with the single-layer configuration, and the column generation algorithm could reduce the problem scale by 92\% at most without losing optimality.

\section{System Model and Problem Formulation}
\label{systemmodel}

Consider a Walker Star constellation consisting of $N$ satellites deployed on $P$ orbital planes. All orbits share the same inclination angle and are uniformly distributed along the equator.\footnote{It should be noted that the proposed routing and computation offloading strategy can be utilized for any constellation configuration once the ISLs and SGLs are known.}  Similar to the Starlink constellation \cite{chaudhry2021laser}, each satellite communicates with four adjacent satellites via laser ISLs, including two neighboring satellites in the same orbital plane and two phase-aligned satellites in adjacent orbital planes. A subset of satellites could also communicate with the ground station via laser satellite-to-ground links (SGLs) when they are within the visible arc of the ground station. Furthermore, similar to \cite{wang2022cdmr}, we discretize the dynamic topology of STNs into several time slots, ensuring that the network topology is unchanged within each slot. Besides, the volume of the raw data captured by satellites is assumed to be known and unchanged within each slot, following a log-normal distribution in the spatial domain \cite{liao2023swarm,cao2023computing}.

The network topology is characterized by a graph $\mathcal{G} = (\mathcal{V},\mathcal{E})$, where $\mathcal{V}$ includes a set of satellites $\mathcal{V}^s$ and the ground station indexed by $0$, and $\mathcal{E}$ is the set of edges including both ISLs and SGLs, denoted by $\mathcal{E}^{s}$ and $\mathcal{E}^{g}$, respectively, i.e., $\mathcal{V} = \mathcal{V}^s \cup \{0\}$ and $\mathcal{E} = \mathcal{E}^{s} \cup \mathcal{E}^{g}$. Moreover, as noted in \cite{mcmenamy2019hop,liao2023swarm}, multi-hop routing brings an additional challenge of the increased latency for every hop. To satisfy the latency requirements, the maximum allowed number of hops needs to be limited for both inter-satellite and satellite-to-ground routes, which are denoted by $H^s$ and $H^g$, respectively. The feasible inter-satellite route is defined as a route sourcing from a satellite and terminating at another satellite while satisfying the hop constraint $H^s$. Similarly, the feasible satellite-to-ground route is defined as a route sourcing from a satellite and terminating at the ground station while satisfying the hop constraint $H^g$. The sets of all feasible inter-satellite and satellite-to-ground routes are denoted by $\mathcal{P}^s$ and $\mathcal{P}^s$, respectively. To formulate the routing and task offloading problem, we introduce two sets of variables. The variables $x^l_i$ represent the volume of raw data collected by the satellite $i$ and computed locally. The variables $f^s_{p^s}$ and $f^g_{p^g}$ denote the traffic volume transmitted by the inter-satellite route $p^s$ and the satellite-to-ground route $p^y$. Furthermore, there are four sets of constraints used to formulate the problem.

\textit{1) Communication capacity constraints:} Firstly, the total data flow routed through a given ISL cannot exceed its communication capacity. Define the subsets $\mathcal{P}^s_{(i,j)} \subseteq \mathcal{P}^s$ and $\mathcal{P}^g_{(i,j)} \subseteq \mathcal{P}^g$ are the feasible inter-satellite and satellite-to-ground routes passing the edge $(i,j)$. The communication capacity constraints for ISLs can be written as,
\begin{equation}
\begin{aligned}
\label{satcom_con}
\sum_{p^s \in \mathcal{P}^s_{(i,j)}} \!\!\!\!\! f^s_{p^s} + \!\!\! \sum_{p^g \in \mathcal{P}^g_{(i,j)}} \!\!\!\!\! f^g_{p^g} \leq R^s_{(i,j)}, \; \forall (i,j) \in \mathcal{E}^s,
\end{aligned} 
\end{equation} 
where $R^s_{(i,j)}$ is the communication capacity of the ISL $(i,j)$. Similarly, the communication constraints for SGLs are as follows,
\begin{equation}
\begin{aligned}
\label{groundcom_con}
\sum_{p^g \in \mathcal{P}^g_{(i,0)}} \!\!\!\!\! f^g_{p^g} \leq R^g_{(i,0)}, \; \forall (i,0) \in \mathcal{E}^g,
\end{aligned} 
\end{equation} 
where $R^g_{(i,0)}$ is the communication capacity of the SGL $(i,0)$. It should be noted that only satellites located in the visible arc of the ground station have SGLs. 

\textit{2) Computation capacity constraints:} Due to the limited computing resources onboard, the computation tasks offloaded to each satellite cannot exceed its capacity. For ease of analysis, we assume that all raw data have the same processing density on all satellites, denoted by $\eta$ and measured by cycles/bit. And the computation capacity of the satellite $i$ is $c_i$ in cycles/sec. Therefore, the computation capacity of the satellite $i$ can be calculated by $C_i \triangleq c_i/\eta$ in bits/sec, and the constraint can be written as \cite{deng2022task},
\begin{equation}
\begin{aligned}
\label{compute_con}
x^l_i + \!\!\! \sum_{p^s \in \mathcal{P}^s_{\text{end:}i}} \!\!\!\!\! f^s_{p^s} \leq C_i, \; \forall i \in \mathcal{V}^s,
\end{aligned} 
\end{equation} 
where $\mathcal{P}^s_{\text{end:}i} \subseteq \mathcal{P}^s$ denotes the set of inter-satellite routes terminating at the satellite $i$. It should be noted that the ground station is regarded as a cloud device with massive computing resources. 

\textit{3) Data volume constraints:} The following constraints ensure that the total data volume processed locally, offloaded to other satellites, and offloaded to the ground station does not exceed the total raw data collected by the satellite,
\begin{equation}
\begin{aligned}
\label{data_con}
x^l_i + \!\!\! \sum_{p^s \in \mathcal{P}^s_{\text{start:}i}} \!\!\!\!\! f^s_{p^s} + \!\!\! \sum_{p^g \in \mathcal{P}^g_{\text{start:}i}} \!\!\!\!\! f^g_{p^g} \leq D_i, \; \forall i \in \mathcal{V}^s,
\end{aligned} 
\end{equation} 
where $\mathcal{P}^s_{\text{start:}i} \subseteq \mathcal{P}^s$ and $\mathcal{P}^g_{\text{start:}i} \subseteq \mathcal{P}^g$ respectively denotes the set of inter-satellite and satellite-to-ground routes sourcing from the satellite $i$, and $D_i$ denotes the volume of raw data gathered by the satellite $i$.

Accordingly, the computation capacity maximization problem can be formulated as follows,
\begin{subequations}
\label{formulation}
\begin{align}
\max_{\{x^l_i\},\{f^s_{p^s}\}, \atop \{f^g_{p^g}\}} & \sum_{i \in \mathcal{V}} ( a x^l_i + b \!\!\!\!\! \sum_{p^s \in \mathcal{P}^s_{\text{start:}i}} \!\!\!\!\! f^s_{p^s} + c \!\!\!\!\! \sum_{p^g \in \mathcal{P}^g_{\text{start:}i}} \!\!\!\!\! f^g_{p^g})  \label{Obj} \\
\rm{s.t.} \;
& \eqref{satcom_con},\, \eqref{groundcom_con}, \,\eqref{compute_con}, \,\eqref{data_con}, \\
&  x^l_i \geq 0, \; \forall i \in \mathcal{V}, \\
&  f^s_{p^s} \geq 0, \; \forall p^s \in \mathcal{P}^s, \\
&  f^g_{p^g} \geq 0, \; \forall p^g \in \mathcal{P}^g, 
\end{align}
\end{subequations}where $a$, $b$ and $c$ represent the predefined weight parameters for computing locally, by other satellites and by the ground station, respectively. It can be observed that the problem \eqref{formulation} is a LP problem that can be solved via interior point method in polynomial time. However, it is infeasible to traverse all feasible routes in $\mathcal{P}^s$ and $\mathcal{P}^g$ because $|\mathcal{P}^s|$ and $|\mathcal{P}^g|$ grows exponentially as $N$, $H^s$ and $H^g$ increase, where $|\cdot|$ denotes the cardinality of a set. In other words, the challenge in solving \eqref{formulation} is that the number of variables $f^s_{p^s}$ and $f^g_{p^g}$ increases exponentially with the number of satellites and the maximum allowed hops. Therefore, the column generation is proposed in the following Section \ref{proposed_method} to overcome the curse of dimensionality. 

\section{Column Generation Approach}
\label{proposed_method}

As mentioned above, the main challenge in solving the LP problem \eqref{formulation} is the extremely large number of variables $\{f^s_{p^s}\}$ and $\{f^g_{p^g}\}$. Column generation is an efficient approach for large-scale LP problems \cite{lubbecke2005selected}, where a small subset of variables is initially activated to solve a reduced problem, and variables that can improve the objective function are incorporated iteratively until no further improvements are possible. Therefore, column generation is employed to achieve the global optimal solution of \eqref{formulation}, 

Initially, we select a small set of feasible inter-satellite and satellite-to-ground routes, and denote these two reduced sets as $ \mathcal{P}^{s'}$ and $\mathcal{P}^{g'}$, i.e., $\mathcal{P}^{s'} \subseteq \mathcal{P}^s$ and ${\mathcal{P}^g}' \subseteq \mathcal{P}^g$. After activating the feasible routes in $\mathcal{P}^{s'}$ and $\mathcal{P}^{g'}$ for \eqref{formulation}, the reduced-scale problem is referred as the restricted master problem and can be formulated as,
\begin{subequations}
\label{reduced_form}
\begin{align}
& \max_{\{x^l_i\},\{f^s_{p^s} | \forall p^s \in \mathcal{P}^{s'}\}, \atop \{f^g_{p^g} | \forall p^g \in \mathcal{P}^{g'}\}} \sum_{i \in \mathcal{V}} ( a x^l_i + b \!\!\!\!\! \sum_{p^s \in \mathcal{P}^{s'}_{\text{start:}i}} \!\!\!\!\! f^s_{p^s} + c \!\!\!\!\! \sum_{p^g \in \mathcal{P}^{g'}_{\text{start:}i}} \!\!\!\!\! f^g_{p^g})  \label{reduced_Obj} \\
\rm{s.t.} \;
& \sum_{p^s \in \mathcal{P}^{s'}_{(i,j)}} \!\!\!\!\! f^s_{p^s} + \!\!\!\sum_{p^g \in \mathcal{P}^{g'}_{(i,j)}} \!\!\!\!\! f^g_{p^g} \leq R^s_{(i,j)}, \; \forall (i,j) \in \mathcal{E}^s, \label{reduced_satcom_con} \\
& \sum_{p^g \in \mathcal{P}^{g'}_{(i,0)}} \!\!\!\!\! f^g_{p^g} \leq R^g_{(i,0)}, \; \forall (i,0) \in \mathcal{E}^g, \label{reduced_groundcom_con} \\
& x^l_i + \!\!\! \sum_{p^s \in \mathcal{P}^{s'}_{\text{end:}i}} \!\!\!\!\! f^s_{p^s} \leq C_i, \; \forall i \in \mathcal{V}^s, \label{reduced_compute_con} \\
& x^l_i + \!\!\! \sum_{p^s \in \mathcal{P}^{s'}_{\text{start:}i}} \!\!\!\!\! f^s_{p^s} + \!\!\! \sum_{p^g \in \mathcal{P}^{g'}_{\text{start:}i}} \!\!\!\!\! f^g_{p^g} \leq D_i, \; \forall i \in \mathcal{V}^s, \label{reduced_data_con} \\
&  x^l_i \geq 0, \; \forall i \in \mathcal{V}, \\
&  f^s_{p^s} \geq 0, \; \forall p^s \in \mathcal{P}^{s'}, \\
&  f^g_{p^g} \geq 0, \; \forall p^g \in \mathcal{P}^{g'}, 
\end{align}
\end{subequations}where $\mathcal{P}^{s'}_{(i,j)}$,  $\mathcal{P}^{s'}_{\text{end:}i}$ and $\mathcal{P}^{s'}_{\text{start:}i}$ are the subsets of $\mathcal{P}^{s'}$ and correspond to the sets $\mathcal{P}^{s}_{(i,j)}$,  $\mathcal{P}^{s}_{\text{end:}:i}$ and $\mathcal{P}^{s}_{\text{start:}i}$ in \eqref{formulation}, respectively. Similarly, $\mathcal{P}^{g'}_{(i,j)}$, $\mathcal{P}^{g'}_{(i,0)}$ and $\mathcal{P}^{g'}_{\text{start:}i}$ are the subsets of $\mathcal{P}^{g'}$ and correspond to the sets $\mathcal{P}^{g}_{(i,j)}$, $\mathcal{P}^{g}_{(i,0)}$ and $\mathcal{P}^{g}_{\text{start:}i}$ in \eqref{formulation}, respectively. It can be observed that \eqref{reduced_form} is a LP problem and the dual problem can be written as,
\begin{subequations}
\label{dual_form}
\begin{align}
\min_{\{\alpha_{ij}\},\{\beta_{i0}\}, \atop \{\gamma_i\},\{\zeta_i\}}  & \sum_{(i,j) \in \mathcal{E}^{s}} \!\!\!\!\! R_{(i,j)} \alpha_{ij} + \sum_{(i,0) \in \mathcal{E}^{g}} \!\!\!\!\! R_{(i,0)} \beta_{i0} \notag \\ & + \sum_{i \in \mathcal{V}} \!\! D_i \gamma_i + \sum_{i \in \mathcal{V}} \!\! C_i \zeta_i \label{dual_Obj} \\
\rm{s.t.} \;
& \sum_{(i,j) \in p^s} \!\!\!\!\! \alpha_{ij} + \gamma_{s_{p^s}} + \zeta_{t_{p^s}} \geq b, \; \forall p \in \mathcal{P}^{s'}, \label{dual_sat_con} \\
& \sum_{(i,j) \in p^g} \!\!\!\!\! \alpha_{ij} + \beta_{t_{p^g}0} + \gamma_{s_{p^g}} \geq c, \; \forall p \in \mathcal{P}^{g'}, \label{dual_ground_con} \\
& \gamma_i + \zeta_i \geq a, \; \forall i \in \mathcal{V}^s, \label{dual_local_con} \\
&  \alpha_{ij} \geq 0, \; \forall (i,j) \in \mathcal{E}^s, \label{dual_alpha_nonneg} \\
&  \beta_{i0} \geq 0, \; \forall (i,0) \in \mathcal{E}^g, \\
&  \gamma_i \geq 0, \; \zeta_i \geq 0, \; \forall i \in \mathcal{V}^s,
\end{align}
\end{subequations}where $\{\alpha_{ij}\}, \; \{\beta_{i0}\}, \; \{\gamma_i\}$ and $ \{\zeta_i\}$ are dual variables corresponding to the constraints \eqref{reduced_satcom_con},\eqref{reduced_groundcom_con},\eqref{reduced_compute_con} and \eqref{reduced_data_con}, respectively. The notation $(i,j) \in p^s$ and $(i,j) \in p^g$ indicate that the routes $p^s$ and $p^g$ passe through the edge $(i,j)$, respectively. The indexes $s_{p^s} \in \mathcal{V}^s$ and $t_{p^s} \in \mathcal{V}^s$ denote the source and terminal node of route $p^s$, respectively, $s_{p^g} \in \mathcal{V}^s$ denotes the source satellite of route $p^g$, and $t_{p^g} \in \mathcal{V}^s$ denotes the satellite in route $p^g$ that is connected to the ground station. Problem \eqref{dual_form} is generally referred as the pricing problem. According to the basic principle of column generation, we have the following Lemma 1.

\textit{\textbf{Lemma 1:}} If \eqref{dual_sat_con} is satisfied for all $p^s \!\in \! \mathcal{P}^s$ and \eqref{dual_ground_con} is satisfied for all $p^g \!\in \! \mathcal{P}^g$ (i.e., not only for $p^s \! \in \! \mathcal{P}^{s'}$ and $p^g \! \in \! \mathcal{P}^{g'}$), the solution of the restricted problem \eqref{reduced_form} is also globally optimal for the original problem \eqref{formulation}.

\textit{Proof:} Denoting the dual problem of the original problem \eqref{formulation} as (\ref{formulation}-D), the optimal solutions of the problem \eqref{formulation}, (\ref{formulation}-D), \eqref{reduced_form}, and \eqref{dual_form} are represented by {\textsf{OPT}}\eqref{formulation}, {\textsf{OPT}}(\ref{formulation}-D), {\textsf{OPT}}\eqref{reduced_form}, and {\textsf{OPT}}\eqref{dual_form}, respectively. 
\begin{align}
\label{opt_relation}
{\textsf{OPT}}\eqref{dual_form} \overset{(a)}{=} {\textsf{OPT}}\eqref{reduced_form} \overset{(b)}{\leq} {\textsf{OPT}}\eqref{formulation} \overset{(c)}{=} {\textsf{OPT}}(\ref{formulation}\textrm{-D}), 
\end{align} 
where (a) and (c) hold because of the strong duality of the LP problem, and (b) holds because $\mathcal{P}^{s'}$ and $\mathcal{P}^{g'}$ are restricted subsets of $\mathcal{P}^{s}$ and $\mathcal{P}^{g}$, respectively. If \eqref{dual_sat_con} is satisfied for all $p^s \!\in \! \mathcal{P}^s$ and \eqref{dual_ground_con} is satisfied for all $p^g \!\in \! \mathcal{P}^g$, we have the following equation because the LP problem (\ref{formulation}\textrm{-D}) only has one optimal objective value,
\begin{align}
\label{opt_eq}
{\textsf{OPT}}\eqref{dual_form} = {\textsf{OPT}}(\ref{formulation}\textrm{-D}), 
\end{align} 
Substituting \eqref{opt_eq} into \eqref{opt_relation}, the equivalence in (b) will be guaranteed. This completes the proof of Lemma 1. $\square$

It can be observed from Lemma 1 that after solving the restricted master problem \eqref{reduced_form}, its objective value can only be further improved if a route $p^s$ violating \eqref{dual_sat_con} or a route $p^g$ violating \eqref{dual_ground_con} is found and added to the restricted sets $\mathcal{P}^{s'}$ or $\mathcal{P}^{g'}$. Otherwise, the global optimal solution of \eqref{formulation} has been achieved. Consequently, the remaining challenge is to find the routes violating \eqref{dual_sat_con} and \eqref{dual_ground_con}. Hereafter, we would illustrate that this challenge can be regarded as the hop-constrained minimum-weight path problem and solved efficiently using the truncated Bellman-Ford algorithm.

We first focus on whether there is a feasible inter-satellite route $p^s$ that violates the constraint \eqref{dual_sat_con}. In other words, the challenge is to find a route $p^s \in \mathcal{P}^{s} \! \setminus \! \mathcal{P}^{s'}$ satisfying the following inequality,
\begin{align}
\label{sat_route_ineq}
\sum_{(i,j) \in p^s} \!\!\!\!\! \alpha_{ij} < b  - \gamma_{s_{p^s}} - \zeta_{t_{p^s}}.
\end{align}
It should be noted that the value of the dual variables $\{\alpha_{ij}\}, \;  \{\gamma_i\}$ and $ \{\zeta_i\}$ can be naturally obtained after solving the restricted master problem \eqref{reduced_form}. To solve this problem, we construct a graph only consisting of the satellites and ISLs, i.e., $\mathcal{G}^s = (\mathcal{V}^s,\mathcal{E}^s)$, and assign the weight value $\alpha_{ij}$ to each corresponding edge $(i,j) \! \in \! \mathcal{E}^s$ and the weights $\gamma_i, \; \zeta_i$ to corresponding vertex $i \! \in \! \mathcal{V}^s $. Afterwards, traverse  all source-destination pairs $\{(s_{p^s}, t_{p^s}) \;|\; s_{p^s} \! \in \! \mathcal{V}^s, \; t_{p^s} \! \in \! \mathcal{V}^s, \; s_{p^s} \neq t_{p^s}\}$ to find the minimum-weight path from $s_{p^s}$ to $t_{p^s}$ that satisfies the maximum-hop constraint $H^s$. If such a path satisfies the inequality \eqref{sat_route_ineq}, it is considered to have the potential to improve the objective value \eqref{reduced_Obj} and should be added to the reduced set $\mathcal{P}^s$. Because the edge weights $\alpha_{ij}$ are nonnegative, the Bellman-Ford algorithm could be modified and utilized to solve the hop-constrained minimum-weight path problem \cite{cormen2022introduction}. Select a source node $s_{p^s} \! \in \! \mathcal{V}^s$, and denote the weight value of the path from $s_{p^s}$ to the vertex $j \! \in \! \mathcal{V}^s$ that consists of no more than $m$ edges by  $u_{s_{p^s} \to j}^{(m)}$. The value can be initialized as follows,
\begin{subequations}
\label{BF_method_ini}
\begin{empheq}[left={\empheqlbrace\,}]{align}
& u_{s_{p^s} \to j}^{(1)} = \alpha_{s_{p^s}j},  \; {\rm if} \; (s_{p^s}, j) \in \mathcal{E}^s, \label{BF_ini1} \\
& u_{s_{p^s} \to j}^{(1)} = \infty, \;\;\;\; {\rm otherwise.} \label{BF_ini2}
\end{empheq}
\end{subequations}Consequently, the value of $u_{s_{p^s} \to j}^{(m)}$ could be updated by,
\begin{align}
\label{BF_update}
u_{s_{p^s} \to j}^{(m+1)} = \min \big\{u_{s_{p^s} \to j}^{(m)}, \, \min_{(i,j) \in \mathcal{E}^s}\{u_{s_{p^s} \to i}^{(m)} + \alpha_{ij}  \}\big\}.
\end{align} 
The update process is terminated when $m+1=H^s$ to ensure that the routes comply with the hop constraint. As a result, the minimum-weight value of the route from $s_{p^s}$ to satellite $j$ is computed and denoted as $u_{s_{p^s} \to j}^{H^s}$. If $u_{s_{p^s} \to j}^{H^s}$ satisfies the inequality \eqref{sat_route_ineq}, the corresponding inter-satellite route is considered to have the potential to increase the objective value in \eqref{reduced_Obj} and is thus included in the set $\mathcal{P}^{s'}$.

% \begin{algorithm}[!t]
% \caption{Maximize STN Capacity via Column Generation}
% \label{CG_alg}
% \begin{algorithmic}[1]
% \STATE Initialize the restricted subsets $\mathcal{P}^{s'} \subseteq \mathcal{P}^s$ and $\mathcal{P}^{g'} \subseteq \mathcal{P}^g$. \label{CG_line1}
% \REPEAT
% \STATE Solve the restricted master problem \eqref{reduced_form}, and obtain the dual variables $\{\alpha_{ij}\}, \; \{\beta_{i0}\}, \; \{\gamma_i\}$ and $ \{\zeta_i\}$. 
% \STATE Assign the value of $\{\alpha_{ij}\}$ to $\mathcal{G}^s$. Traverse all satellites as the source node and solve the minimum-weight path problem via the truncated Bellman-Ford algorithm.
% \IF{There is a route $p^s$ satisfy \eqref{sat_route_ineq}}
% \STATE Add $p^s$ to $\mathcal{P}^{s'}$.
% \ENDIF
% \STATE Assign the value of $\{\alpha_{ij}\}$ and $\{\beta_{i0}\}$ to $\mathcal{G}$. Traverse all satellites as the source node and solve the minimum-weight path problem.
% \IF{There is a route $p^g$ satisfy \eqref{ground_route_ineq}}
% \STATE Add $p^g$ to $\mathcal{P}^{g'}$.
% \ENDIF
% \UNTIL{No route is added to $\mathcal{P}^{s'}$ and $\mathcal{P}^{g'}$ in this iteration.}
% \STATE The optimal solution of \eqref{reduced_form} is globally optimal for \eqref{formulation}.
% \end{algorithmic}
% \end{algorithm}

Similarly, the minimum-weight of satellite-to-ground routes $p^g \in \mathcal{P}^{g} \!\! \setminus \!\! \mathcal{P}^{g'}$ could be solved via the truncated Bellman-Ford algorithm. By checking if the minimum weigh violate the dual constraint \eqref{dual_ground_con}, i.e., satisfy the following inequality,
\begin{align}
\label{ground_route_ineq}
\sum_{(i,j) \in p^g} \!\!\!\!\! \alpha_{ij} + \beta_{t_{p^g}0} <  c - \gamma_{s_{p^g}},
\end{align}
we can select the feasible satellite-to-ground routes that could improve the the objective value in \eqref{reduced_Obj} and add them to the set $\mathcal{P}^{g'}$. The column generation algorithm terminates when no routes  $p^s$violating \eqref{dual_sat_con} and $p^g$ violating \eqref{dual_ground_con} can be identified. Moreover, a realization of column generation is open-sourced in \href{https://github.com/Yuanliaoo/CG_satellite}{github.com/Yuanliaoo/CG\_satellite}.

\textit{\textbf{Remark 1:}} (Computation complexity) Recalling that each iteration of the column generation algorithm involves solving the restricted master problem problem \eqref{reduced_form} and the hop-constrained minimum-weight path problem, the complexity of these two steps would be analyzed separately. First, in the section 6.6.1 of \cite{ben2001lectures}, the complexity of solving linear programming is given by $\mathcal{O}\big((n^v+n^c)^{1.5}{n^v}^2\big)$, where $n^v$ denotes the number of variables and is equal to $ |\{x^l_i\}|+|\{\mathcal{P}^{s'}\}| + |\{\mathcal{P}^{g'}\}|$ in \eqref{reduced_form}, $n^c$ represents the number of constraints and is equal to $|\mathcal{E}^s| + |\mathcal{E}^g| + 2|\mathcal{V}^s|$ in \eqref{reduced_form}. Second, the truncated Bellman-Ford algorithm is applied to calculate the minimum-weight path sourced from all satellites, the computation complexity of which is $\mathcal{O}\big(|\mathcal{V}^s| (H^s|\mathcal{E}^s|+H^g|\mathcal{E}^g|) \big)$, as detailed in \cite{cormen2022introduction}. It is evident that the overall computational complexity is primarily dominated by $n^v$ when solving \eqref{reduced_form}. Furthermore, the following Table \ref{scale_reducation} in Section \ref{NumericalResults} demonstrates that the column generation algorithm effectively reduces the value of $n^v$. 
%Consequently, the proposed column generation algorithm offers an efficient approach for solving the original problem \eqref{formulation}.

\section{Numerical Investigations}
\label{NumericalResults}

\begin{figure}[!t]
\centering
\includegraphics[width=.675\linewidth]{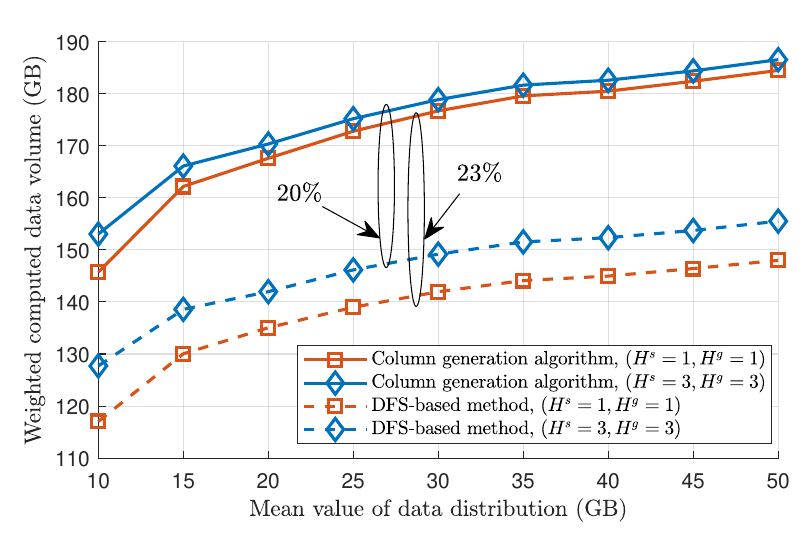}
\caption{Computed data volume under different data distribution.}
\label{diff_mean}
\end{figure}

In this section, numerical results are presented to evaluate the proposed three-layer MEC protocol. The simulations are based on a Walker Star constellation consisting of 30 satellites distributed on 6 orbital planes, with 6 satellites linked to the ground station via SGLs. The simulation parameters are listed in Table \ref{Notation}. Furthermore, similar to \cite{cao2023computing, wang2015approach}, it is assumed that the volume of raw data generated by each satellite for processing follows the lognormal distribution, with a mean of 20 GB (except in Fig. \ref{diff_mean}) and a standard deviation of 1.3. 

\begin{table}[!t]
\centering
\caption{Summary Of Notations}
\label{Notation}
\begin{tabular}{p{5.2 cm}|l}
\hline
\textbf{Parameters} & \textbf{Value}\\
\hline
Communication capacity of the ISLs $R^s_{(i,j)}$ & 5 Gbps \cite{cao2023computing}\\
Communication capacity of the SGLs $R^g_{(i,0)}$ & 1 Gbps \cite{cao2023computing} \\
Computation capacity for satellites $c_i$ & $10^9$ cycles/sec \\
Processing density $\eta$ & $10^8$ cycles/Gbits \\
Weight parameters [a, b, c] & [0.6, 0.3, 0.1] \\
\hline
\end{tabular}
\end{table}

Fig. \ref{diff_mean} illustrates the value of the objective function \eqref{Obj} versus different raw data distributions solved by different methods. Recalling that the objective function \eqref{Obj} calculates the weighted sum of the computed data volume, Fig. \ref{diff_mean} shows that the value of \eqref{Obj} first increases and then levels off as the mean value increases. This is because the computing capacity of the STN, limited by the computing and communication resources, is gradually approached. Furthermore, we employ the depth-first search (DFS) based method as a benchmark, i.e., each satellite prioritizes local computation and sequentially offloads residual tasks to neighboring nodes if local capacity is exceeded, adhering to maximum-hop constraints, as well as the communication and computation capacity limits. The DFS-based method propagates tasks recursively to subsequent nodes until all tasks are processed or constraints are violated. It can be observed in Fig. \ref{diff_mean} that the proposed column generation algorithm outperforms the DFS-based method in terms of the objective value. Numerically, the column generation algorithm improves the weighted computed data volume by 23\% and 20\% when setting $H^s \! = \! H^g  \! = \! 1$ and $H^s \! = \! H^g \! = \! 3$, respectively. 

\begin{figure}[!t]
\centering
\includegraphics[width=.68\linewidth]{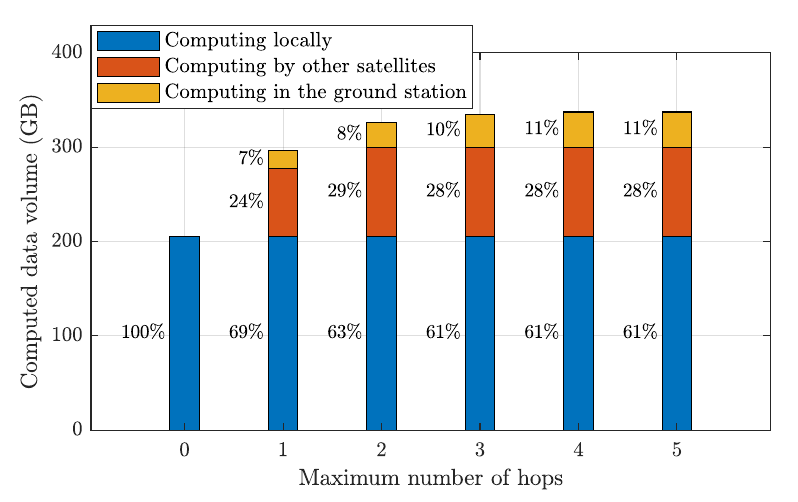}
\caption{Data volume computed on different layers.}
\label{diff_hops}
\end{figure}

Fig. \ref{diff_hops} presents the computed data volume versus different maximum allowable hops, where we assume $H^s = H^g$ for simplicity. It can be observed that the computation capacity of the STN increases as the maximum number of hops grows when $H^s \! =  \! H^g \! \leq  \! 4$. The reason is that larger $H^s$ and $H^g$ allow a wider transmission of raw data, enabling computation on nodes with surplus computing resources. The growth in computation capacity reaches a plateau when the maximum number of hops is greater than 4 due to the communication constraints of the ISLs and SGLs. Notably, when the maximum-hop is constrained to 0, the three-layer MEC protocol degenerates into a single-layer configuration that performs only local computation. As illustrated in Fig. \ref{diff_hops}, the three-layer protocol improves the computation capacity by 60\% when setting $H^s \! = \! H^g \! = \! 5$. Furthermore, observing the data volume computed locally, on other satellites, and at the ground station, it can be seen that the improvement in computation capacity is driven by the increasing volume of data processed on other satellites and at the ground station.

Recalling that one of the main contributions of this letter is employing the column generation algorithm to overcome the exponential growth in the problem scale of \eqref{formulation}. As mentioned in Section \ref{proposed_method}, the column generation method could achieve the optimal solution of \eqref{formulation} by activating only a subset of variables. Table \ref{scale_reducation} illustrates the significant reduction in problem scale numerically. For example, when setting $H^s=H^g=5$, the original problem contains 13938 variables to denote the data flow transmitting in feasible routes. However, by using the column generation algorithm, only 1080 variables are activated to achieve the optimal solution, i.e., the problem scale is reduced by 92\%. 
% Therefore, it is shown that the column generation algorithm is an efficient approach for solving large-scale routing and computation offloading problems in STNs.

\section{Conclusions}
\label{Conclusion}

In this letter, a three-layer edge computing protocol is proposed to enhance the computation capacity in satellite-terrestrial networks (STNs), in which the computing-intensive tasks can either be processed locally, or transmitted to other satellites or the ground station for processing. To this end, the routing planning and task offloading problem is formulated as a linear programming (LP) problem and solved by column generation to overcome the exponential growth in problem scale. Numerical investigations reveal that the proposed three-layer computing protocol can enhance the computing capacity by 60\% compared to the single-layer configuration that only allows for local processing, and the proposed column generation algorithm could reduce the problem scale by 92\% at most without losing optimality.

\begin{table}[!t]
\centering
\caption{Problem scale reduced by column generation}
\label{scale_reducation}
\begin{tabular}{|p{0.5 cm}|p{0.5 cm}|p{1.7 cm}|p{1.7 cm}|p{2 cm}|}
\hline
$H^s$ & $H^g$ &\textbf{$|\mathcal{P}^s| + |\mathcal{P}^g|$} & \textbf{$|\mathcal{P}^{s'}| + |\mathcal{P}^{g'}|$} & \textbf{Scale reduction} \\
\hline
1 & 1 & 126 & 126 & 0\% \\
2 & 2 & 510 & 379 & 26\% \\
3 & 3 & 1662 & 666 & 60\%\\
4 & 4 & 4878 & 852 & 83\%\\
5 & 5 & 13938 & 1080 & 92\%\\
\hline
\end{tabular}
\end{table}

\bibliographystyle{IEEEtran}
\bibliography{IEEEabrv,reference} 

\end{document}